\documentstyle[twocolumn,aps,epsfig]{revtex}
\begin{document}
\draft
\title{The optical
conductivity of the quasi one-dimensional organic
conductors: the role of forward scattering by impurities}
\author{Peter Kopietz$^{1}$ and Guillermo E. Castilla$^{2}$} 
\address{
$^{1}$Institut f\"{u}r Theoretische Physik, Universit\"{a}t G\"{o}ttingen,
Bunsenstrasse 9, D-37073 G\"{o}ttingen, Germany\\
$^{2}$Department of Physics, University of California, Riverside,
California 92521}
\date{December 17, 1998}
\maketitle
\begin{abstract}
We calculate the average 
conductivity $\sigma ( \omega )$ of 
interacting electrons in one dimension in the presence of
a long-range random potential (forward scattering disorder). 
Taking the curvature of the energy dispersion into account,
we show that weak disorder  
leads to 
a transport scattering rate 
that vanishes as $\omega^2$ for small frequency
$\omega$. 
This implies ${\rm Im} \sigma ( \omega ) \sim \tilde{D}_c / \omega$ 
and ${\rm Re} \sigma ( \omega ) \sim  \tilde{D}_c \tau $ for
$\omega \rightarrow 0$, where $\tilde{D}_c$ is the 
renormalized charge stiffness and 
the time $\tau$ is proportional to the strength
of the impurity potential.
These non-trivial effects due to forward scattering
disorder are lost within the usual bosonization
approach, which relies on the linearization of
the energy dispersion.
We discuss our result in the light of a recent
experiment.

\end{abstract}
\pacs{PACS numbers: 78.20.-e, 71.10.Pm, 75.30.Fv}
\narrowtext
%
%
%
In a recent measurement of the
frequency-dependent conductivity $\sigma ( \omega)$
of quasi one-dimensional organic conductors
of the (TMTSF)$_2$X-series,
Schwartz {\it et al.}\cite{Schwartz98} found that
for small frequencies $\omega$
the data could be fitted by
 \begin{equation} 
 \sigma ( \omega ) =  \frac{D_c}{  
 \Gamma ( \omega ) - i \omega \frac{m^{\ast} ( \omega) }{ m_b} }
 \; ,
 \label{eq:sigmaform}
 \end{equation}
where $D_c$ is the bare charge stiffness, 
and where
the transport scattering rate $\Gamma ( \omega )$ and
the effective mass enhancement
$m^{\ast} ( \omega ) / m_b$  
are given by
 \begin{equation}
 \Gamma ( \omega ) = \Gamma_0  +
 \frac{ \lambda_0 \alpha \omega^2}{
 1 + \alpha^2 \omega^2 }
 \label{eq:Gamma}
 \; ,
 \end{equation}
 \begin{equation}
 \frac{m^{\ast} ( \omega ) }{ m_b} = 1 + 
 \frac{ \lambda_0}{
 1 + \alpha^2 \omega^2 }
 \; .
 \label{eq:meff}
 \end{equation}
The quadratic frequency-dependence in Eqs.(\ref{eq:Gamma}) and
(\ref{eq:meff}) is characteristic of Fermi
liquids in three dimensions.
Expressions
of this type were used by Sulewski {\it{et al.}}\cite{Sulewski88}
in their study of the 
compound UPt$_3$, which is a three-dimensional Fermi liquid 
with heavy mass.
The experimentally obtained values\cite{Schwartz98,Throughout}
for a sample consisting of (TMTSF)$_2$PF$_6$ are
$\Gamma_0 / (2 \pi c )= 0.56 cm^{-1}$, $\lambda_0 = 1$,
and $1 / ( 2 \pi c \alpha ) = 1 cm^{-1}$.
Schwartz {\it{et al.}}\cite{Schwartz98}
speculated
that the physical origin of the quadratic frequency-dependence
of the second term in Eq.(\ref{eq:Gamma})
is inelastic electron-electron scattering
in a clean three-dimensional Fermi liquid.
The physical picture is that at
sufficiently low temperatures, there is a 
crossover from one-dimensional 
behavior at higher energies to
a regime characterized by three-dimensional 
phase coherence: this is a consequence of the
finite coupling between the chains\cite{Comment}.
Experimental evidence for the existence of such a crossover 
in the organic conductors has
also been given by Moser {\it{et al.}}\cite{Moser98}
via DC transport measurements.
However, 
the authors of Ref.\cite{Schwartz98} emphasized that
the
interpretation of the
frequency-dependence in Eqs.(\ref{eq:Gamma},\ref{eq:meff})
in terms of Fermi liquid theory
leads to an anomalously small value of $1/ \alpha$. 
Moreover, the Fermi surface of the organic conductors
is nested, so that one should expect a scattering rate
linear in $\omega$\cite{Ruvalds91},
in disagreement with the experimental result\cite{Schwartz98}.
In this note we would like to point
out that there exists an alternative 
(and in our opinion physically more plausible)
explanation
for the quadratic frequency-dependence of the scattering rate
and the finite effective mass renormalization
in the organic conductors,
namely {\it{forward scattering
by impurities}}.
As we shall show below, this interpretation of the data
also leads to a natural explanation
of the anomalously small value of $1/ \alpha$ seen
in the experiment\cite{Schwartz98}.

As discussed in Ref.\cite{Gorkov85}, in the
quasi one-dimensional organic conductors it is natural to
expect that the disorder potential seen by the
electrons on the chains is weak and slowly varying
along the chain. Such a potential can be modelled by a 
Gaussian random potential $U ( x )$ with zero average
and long-range correlator
 \begin{equation}
 \overline{ U (x) U ( x^{\prime})} = \gamma_0 
 C (    x - x^{\prime}  ) 
 \label{eq:correlator}
 \; ,
 \end{equation}
where the overline denotes averaging over the disorder.
Here $\gamma_0$ is a measure of the strength
of the disorder, 
and $C (  x )$ is assumed to 
have a maximal range $\xi$ that is large compared with
$(2 k_F)^{-1}$,
where $k_F$ is the Fermi wave-vector.
In other words, we assume that 
$C ( x )$ is a finite positive constant
for
$| x | { \raisebox{-0.5ex}{$\; \stackrel{<}{\sim} \;$}} \xi$,
and vanishes for $|x| \gg \xi$.
For convenience we normalize $C ( x )$ such that
its Fourier transform
 \begin{equation}
\tilde{C} ( q)
= \int_{- \infty}^{\infty} dx e^{ i q x } C ( x )
 \end{equation}
is dimensionless. The above properties of $C ( x)$
imply that $  \tilde{C} ( q )$ vanishes 
if $|q| \xi \gg 1$.
The inverse of $\xi$
can be identified with the maximal possible momentum transfer
between two electrons due to scattering by the
impurity potential. The
requirement $\xi^{-1} \ll 2 k_F$ means that
impurities do not give rise to backward scattering,
{\it i.e.}, the random potential is dominated by the
 forward scattering.
If this problem is treated by means of the usual bosonization
approach (with linearized energy dispersion), 
one finds that forward scattering disorder
{\it{does not affect the conductivity at 
all}}\cite{Abrikosov78,Giamarchi88}.
It is also possible to confirm this result by
directly expanding the 
average conductivity in powers of the
impurity potential.  
{\it{For linearized energy dispersion}}
one easily verifies that
all impurity corrections cancel.
This is a consequence of the
closed-loop theorem\cite{Bohr81,Dzyaloshinskii74}, which is
well known in the context of the 
Tomonaga-Luttinger model\cite{Tomonaga50}.
The closed loop theorem
implies that at long wave-lengths 
all closed fermion loops with more than two external
legs vanish after symmetrization\cite{Bohr81,Dzyaloshinskii74}.
For this cancelation to take place, it is irrelevant
whether the external legs represent 
the dynamic Coulomb interaction or static impurity lines.
However, the closed-loop theorem
is {\it only} valid if the
energy dispersion $\epsilon ( k )$ is linearized
close to the Fermi points $\pm k_F$,
which amounts to
ignoring the quadratic and higher terms in the expansion
 \begin{equation}
 \epsilon ( \pm k_F + q ) -  \epsilon ( \pm k_F) =
 \pm v_F q + \frac{ q^2}{2 m_b } + O (q^3) 
 \label{eq:epsexp}
  \; .
  \end{equation}
Here $v_F$ is the Fermi velocity
and $m_b$ is the band mass.
Clearly, in order to calculate the leading effect of 
forward scattering disorder
on the conductivity 
it is insufficient to work with linearized energy dispersion,
as it is done usually in the bosonization approach.

We now present a simple calculation of the
effect of forward scattering disorder
on the conductivity
of an interacting electron gas in one dimension.
We assume that
the electron gas is metallic and
focus 
on energy scales smaller than possible
spin gaps, so that 
we can ignore backward and umklapp
scattering. 
In principle, one could try to treat 
the non-linear terms in the energy dispersion
within the framework of bosonization, 
as it was recently done
in a different context in Ref.\cite{Fazio97}.
However, there  is a much simpler
and physically more transparent
solution to our problem.
According to G\"{o}tze and W\"{o}lfle\cite{Goetze72},
in the perturbative calculation of the
conductivity it is often useful to 
introduce the memory function $M( \omega )$ 
by setting
 \begin{equation}
 \sigma ( \omega ) =  
 \frac{i D_c }{\omega + M ( \omega )}
  \; ,
 \end{equation}
and calculating $M ( \omega )$ instead of $\sigma ( \omega )$
in powers of the impurity 
potential. In this way one implicitly takes into account  
vertex corrections to all orders in perturbation theory.
To leading order in the strength of the impurity
potential one finds in one dimension\cite{Goetze72}
 \begin{eqnarray}
 M (\omega ) & = & \frac{\gamma_0}{n m_b} \int_{ - \infty}^{\infty}
  \frac{d q}{2 \pi} q^2  \tilde{C} ( q ) 
  \nonumber
   \\
   & \times & 
   \left[ \frac{ \Pi ( q , \omega + i 0^{+}) 
  - \Pi ( q , i 0^{+} ) }{\omega + i 0^{+} } \right]
  \; ,
   \label{eq:Mem}
  \end{eqnarray}
where $\Pi ( q , \omega )$ is the 
density-density correlation function and
$n$ is the density of the one-dimensional electron gas.
The memory function approach has also been 
used by Giamarchi to study
the effect of umklapp scattering
on the conductivity of one-dimensional
interacting fermions\cite{Giamarchi91}.
For the Tomonaga-Luttinger model with short-range disorder
Eq.(\ref{eq:Mem}) has been evaluated by
Luther and Peschel\cite{Luther74}.
In this case the anomalous scaling of $\Pi ( q , \omega )$ for
momenta close to $2 k_F$ dominates the conductivity.
In contrast, 
in our case we need 
$\Pi ( q , \omega )$ only  for small momenta, $| q | 
{ \raisebox{-0.5ex}{$\; \stackrel{<}{\sim} \;$}} \xi^{-1}
\ll 2 k_F$. Observe that
the leading effect
of the non-linearity in the energy dispersion
is already contained in the prefactor $1/m_b$ 
in Eq.(\ref{eq:Mem}); thus, we can 
calculate $\Pi ( q , \omega )$
on the right-hand side
of Eq.(\ref{eq:Mem})
for linearized energy dispersion.
For simplicity we substitute for
$\Pi ( q , \omega )$ 
the density-density correlation function of the
Tomonaga-Luttinger model\cite{Tomonaga50} with
interaction parameters $g_2 = g_4 = \pi v_F F$,
 \begin{equation}
 \Pi ( q , \omega ) = Z_q \frac{ 2 \omega_q}{ \omega_q^2 - \omega^2 }
 \; ,
 \label{eq:Pi1}
  \end{equation}
 \begin{equation}
 Z_q = \frac{ | q |}{ \pi \sqrt{ 1 + F } }
 \; \; , \; \;
  \omega_q = \tilde{v}_F | q | 
  \; ,
  \label{eq:Zomega}
  \end{equation}
where $\tilde{v}_F = \sqrt{1 + F} v_F$. 
Using Eq.(\ref{eq:Mem}) and assuming
$ \tilde{C} ( q ) = \Theta ( 1- | q | \xi )$,
we obtain after a simple calculation for 
$| \omega | \ll 
\tilde{v}_F / \xi $
 \begin{eqnarray}
  {\rm Im} M ( \omega + i 0^{+})
  & = &  a \omega^2
  \; ,
  \label{eq:ImM}
  \\
  {\rm Re} M ( \omega  + i 0^{+} ) 
  & = & b \omega  [ 1 + O ( \omega^2 )]
  \label{eq:ReM}
  \; ,
  \end{eqnarray}
where
 \begin{eqnarray}
  a & = &  \frac{ \gamma_0 }{   
  \pi n m_b \tilde{v}_F^4 \sqrt{1 + F} } 
  \label{eq:ares}
  \; ,
  \\
  b & = & 
  \frac{ 2 \gamma_0 }{   \pi^2 n m_b \tilde{v}_F^3 \xi 
  \sqrt{ 1 + F }   }
  \label{eq:bres}
  \; .
  \end{eqnarray}
Note that $a$ and $b$ both vanish for $1/m_b \rightarrow 0$,
corresponding to the linearization of the energy 
dispersion (\ref{eq:epsexp}).
We conclude that
forward scattering disorder leads to a conductivity
of the form given in Eq.(\ref{eq:sigmaform}), with
 \begin{eqnarray}
 \Gamma ( \omega ) & = & a \omega^2
  \;  , 
  \label{eq:Gammares}
   \\
  \frac{m^{\ast} ( \omega )}{m_b}
  & = & 1 + b + O ( \omega^2 )
  \; .
  \label{eq:disres}
  \end{eqnarray}
In the absence of other scattering mechanisms
this implies for the real- and imaginary part of
the conductivity at frequencies $| \omega | \ll \tilde{v}_F / \xi$,
 \begin{eqnarray}
  {\rm Re} \sigma ( \omega ) & =  &
  {\rm Re}
  \left[ \frac{D_c}{ a \omega^2 - i (1+b ) \omega } \right]
  = \frac{\tilde{D}_c \tau }{ 1 + ( \omega \tau )^2 }
  \; ,
  \label{eq:Drudebroad}
  \\
  {\rm Im} \sigma ( \omega ) & =  &
  \frac{ \tilde{D}_c }{\omega [ 1 + ( \omega \tau )^2 ]}
  \label{eq:Drudeim}
  \; ,
  \end{eqnarray}
where the 
renormalized charge stiffness $\tilde{D}_c$ and the
time $\tau$ are given by
 \begin{equation}
 \tilde{D}_c = \frac{D_c}{ 1 + b }
 \; \; , \; \; 
 \tau = \frac{a}{1+b} 
  \; .
  \label{eq:Dren}
  \end{equation}
Keeping in mind that
for the derivation of Eqs.(\ref{eq:Drudebroad},\ref{eq:Drudeim})
we have assumed $| \omega | \ll \tilde{v}_F/ \xi$,
it is easy to see that
in the regime where Eqs.(\ref{eq:Drudebroad},\ref{eq:Drudeim})
are valid the parameter
$ | \omega | \tau  $ is always smaller than unity.
Hence, for $| \omega | \ll \tilde{v}_F / \xi$ we may approximate
${\rm Re} \sigma ( \omega ) \approx \tilde{D}_c \tau$
and
${\rm Im} \sigma ( \omega ) \approx \tilde{D}_c / \omega$.
Note that the imaginary part of the conductivity
exhibits at small frequencies the usual $1 / \omega$ behavior
of a clean system (but with 
renormalized charge stiffness),
whereas  the Drude peak $\pi D_c \delta ( \omega )$
in the real part of the conductivity of the clean system
is completely destroyed by the disorder, and
is replaced by a constant $\tilde{D}_c \tau$.
In this respect
the effect of forward scattering disorder
in one dimension is similar to the effect of weak short-range disorder
in three dimensions.
Note, however,
that according to Eqs.(\ref{eq:ares},\ref{eq:Dren})
the effective lifetime $\tau$ is
proportional to the strength $\gamma_0$ of the impurity potential.
In contrast, for short-range impurity scattering
one finds within the Born approximation that the
{\it{inverse}} lifetime is proportional
to the impurity strength. 
These non-trivial effects associated with long-range disorder
in one dimension
are missed within the usual bosonization approach, which
predicts a 
Drude peak $\pi D_c \delta ( \omega )$
in the real part of the conductivity, even in 
the presence of long-range
disorder\cite{Abrikosov78,Giamarchi88}.

Keeping in mind that Eqs.(\ref{eq:ImM}) and (\ref{eq:ReM})
are the leading terms for small $\omega$, 
Eqs.(\ref{eq:Gammares}) and (\ref{eq:disres}) 
are consistent with the  
frequency-dependent part of the
experimentally seen behavior given
in Eqs.(\ref{eq:Gamma}) and (\ref{eq:meff})\cite{footnote}. 
The constant part $\Gamma_0$
of the scattering rate in Eq.(\ref{eq:Gamma})  
must be due to 
some other (short-range) impurity scattering mechanism
which we have not taken into account in our calculation.
A similar assumption concerning the origin of $\Gamma_0$ 
must be made if one
interprets the frequency-dependence of $\Gamma ( \omega )$
to be due to inelastic 
scattering in a three-dimensional Fermi liquid\cite{Schwartz98}.
Obviously we may identify
$\lambda_0 = b$, and 
 \begin{equation}
 \frac{1}{\alpha} = \frac{b }{ a} = \frac{2 \tilde{v}_F }{\pi \xi} 
 \label{eq:alphares}
  \; .
 \end{equation}
Note that $\alpha$ is independent of $\gamma_0$ and
$m_b$; it depends only on the renormalized
Fermi velocity $\tilde{v}_F$ and 
on the correlation length $\xi$ of the
impurities.
Using the experimental value\cite{Schwartz98}
$ 2 \pi c \alpha  = 1 cm$ 
we obtain from Eq.(\ref{eq:alphares})
$\xi =  \pi^{-2} ( \tilde{v}_F / c ) cm$.
Given a bare interchain hopping $t_{\|} \approx 250 meV$\cite{Schwartz98},
a reasonable estimate for the 
Fermi velocity is
$ \tilde{v}_F \approx 10^{7} cm /s $. 
Then we obtain for the impurity correlation length
$\xi \approx 3 \times 10^{-5} cm$.
This is certainly larger than
$(2 k_F)^{-1}$, so that our interpretation 
of the low-frequency
behavior of the conductivity data\cite{Schwartz98}
in terms of long-range disorder
is internally consistent.
The large value of $\xi$ explains 
the anomalously small value of $\alpha^{-1}$ seen
in the experiment\cite{Schwartz98}.
The impurity correlation length $\xi$
should not be confused with the transport mean free path
$\ell_{\rm tr}$, although 
in the experiment \cite{Schwartz98} both length scales have the same
order of magnitude.
A rough estimate for $\ell_{\rm tr}$ can be
obtained from
Eqs.(\ref{eq:sigmaform},\ref{eq:Gamma}) by identifying
$\ell_{\rm tr} = \tilde{v}_F / \Gamma_0$.
Using the measured value
$\Gamma_0 \approx 2 \pi c \times 0.56 cm^{-1}$, we obtain 
$\ell_{\rm tr}  \approx 9  \times 10^{-5} cm$. 
The importance of long-range disorder in
the organic conductors has been pointed out by
Gorkov\cite{Gorkov85} long time ago:
because defects damage only a small
fraction of the big planar molecules,
it is natural to expect that
the imperfection potential of the defects, as seen
by the electrons on the conducting chains,
is slowly varying along the chains.

An inelastic {\it{single-particle}} scattering rate 
$\Gamma_{1 } (\omega ) \propto \omega^2$
due to electron-electron interactions
is one of the hallmarks of a three-dimensional
Fermi liquid. However, the transport scattering
rate $\Gamma ( \omega )$ in Eq.(\ref{eq:Gammares}) is defined
in terms of a two-particle Green's function,
and it is obviously not related to 
$\Gamma_1 ( \omega )$
of a three-dimensional Fermi liquid.
This shows that  
it is crucial to distinguish between
transport and single-particle scattering rates
in the interpretation of
conductivity measurements of quasi one-dimensional
conductors. 
To understand why 
the $\omega^2$-behavior of the transport scattering
rate $\Gamma ( \omega )$ 
does not imply a similar behavior of the
single-particle scattering rate $\Gamma_1 ( \omega )$, 
one should keep in mind that
the above mentioned cancelations
between self-energy and vertex corrections, which are
responsible for the $\omega^2$-correction 
to the transport scattering
rate, do not occur in the calculation of the
single-particle Green's function.  
In fact, for non-Fermi liquids in arbitrary dimensions
there is no reason to expect that the
transport and single-particle scattering rates are
simply related\cite{Bartosch98}.
Although Eqs.(\ref{eq:Gammares}) and (\ref{eq:disres})
look like conventional Fermi liquid behavior, 
the single-particle Green's function 
of the system exhibits Luttinger liquid behavior. 
For example, the single-particle density of states
vanishes with a non-universal power-law\cite{Bohr81,Tomonaga50}.

So far we have assumed that the system is strictly one-dimensional.
However, realistic experimental systems have a finite hopping $t_{\bot}$ between
the chains, so that it is important to estimate 
the modification of the above result due to a finite value
of $t_{\bot}$. For simplicity, let us assume that the interchain
hopping $t_y = t_{\bot}$ in one direction transverse to the chains
(the $y$-direction)
is much larger than in the other transverse direction.
To a first approximation we may then
ignore the hopping $t_z$ in the $z$-direction.
In view of the anisotropy of the
interchain hopping in the organic conductors\cite{Jerome91}, 
this approximation is not unreasonable.
The transport scattering rate due to forward scattering by impurities
can then be estimated from the two-dimensional 
analog of Eq.(\ref{eq:Mem}), 
 \begin{eqnarray}
 M (\omega ) & = & \frac{\gamma_0}{n_2  m_b} \int_{ - \infty}^{\infty}
  \frac{d q_x d q_y}{ (2 \pi)^2 } q_x^2
  \tilde{C} ( {\bf{q}}  ) 
  \nonumber
   \\
   & \times & 
   \left[ \frac{ \Pi ( {\bf{q}}  , \omega + i 0^{+}) 
  - \Pi ( {\bf{q}}  , i 0^{+} ) }{\omega + i 0^{+} } \right]
  \; ,
   \label{eq:Mem2}
  \end{eqnarray}
where $n_2$ is the two-dimensional density, 
$m_b$ is the effective band mass in the $x$-direction, 
$\tilde{C} ( {\bf{q}} )$ is the two-dimensional
Fourier transform of the disorder correlator,
and $\Pi ( {\bf{q}} , \omega )$ is the two-dimensional 
density-density correlation function.
Keeping in mind that by assumption $\tilde{C} ( {\bf{q}} )$
is dominated by small wave-vectors,
we may use the random-phase approximation to calculate the
density-density correlation function. 
For small $t_{\bot}$ it can be shown\cite{Kopietz97}
that $\Pi ( {\bf{q}} , \omega )$ is dominated by the
plasmon pole, so that 
$\Pi ( {\bf{q}} , \omega )$ 
can be written as in
Eq.(\ref{eq:Pi1}). 
Defining the dimensionless parameter
 \begin{equation}
 \theta = \frac{  | t_{\bot} |}{E_F}
 \; ,
 \end{equation}
where $E_F $ is the Fermi energy,
the residue and the plasmon dispersion 
are for small $\theta$ given by\cite{Kopietz97}
 \begin{equation}
 Z_{\bf{q}} = \frac{ \nu_2 v_F \sqrt{ q_x^2 + \theta^2 q_y^2 }}{ 2 \sqrt{ 1 + F } }
 \; \; , \; \;
  \omega_{\bf{q}} = \tilde{v}_F \sqrt{  q_x^2 + \theta^2 q_y^2}  
  \; ,
  \label{eq:Zomega2}
  \end{equation}
where $\nu_2$ is the two-dimensional density of states.
Assuming for simplicity
$\tilde{C} ( {\bf{q}} ) = \Theta ( 1 - | q_x | \xi )
\Theta ( 1 - | q_y | \xi )$, the low-frequency behavior
of $M ( \omega )$ is easily calculated.
We find that the strictly one-dimensional result 
for $M ( \omega )$
given in Eqs.(\ref{eq:ImM},\ref{eq:ReM}) remains valid as long as
 \begin{equation}
 | \omega | 
{ \raisebox{-0.5ex}{$\; \stackrel{>}{\sim} \;$}} 
 \frac{ \tilde{v}_F  \theta   }{\xi}
 \equiv \omega_{\bot}
 \label{eq:omega2d}
 \; .
 \end{equation}
For $ | \omega | \ll \omega_{\bot} $ 
we find that 
Eqs.(\ref{eq:ImM},\ref{eq:ReM})
should be  multiplied by an extra factor 
of order
$| \omega | / \omega_{\bot}  \ll 1$. Hence, 
the finiteness of the interchain hopping 
induces a crossover 
in the transport scattering rate due to forward scattering
by disorder from a $\omega^2$-behavior
at higher frequencies to a $|\omega|^3$-law at
frequencies $| \omega | \ll \omega_{\bot}$. 
Note, however, that 
the crossover frequency  $\omega_{\bot}$
is of order $ | t_{\bot} | / ( k_F \xi )$, 
so that
for large $\xi$ and small $t_{\bot}$ 
the regime where the $| \omega |^3$-behavior 
is visible is very small and
probably experimentally irrelevant.

Some cautionary remarks are in order:
In this work we have assumed that the scattering due to the
impurities is forward.
It is clear, however, that in the organic conductors
also backward scattering by impurities is present which, 
in a purely one-dimensional system, should lead to localization.
Because this implies a vanishing conductivity,
at sufficiently small frequencies  $\rm Re \sigma ( \omega )$
should become smaller than $\tilde{D}_c \tau$ and
eventually vanish.
However, in the experimental systems  the finite interchain coupling
might stabilize a metallic phase.
Moreover, from the above calculation
it is clear that the constant
part $\Gamma_0$ of the scattering rate in Eq.(\ref{eq:Gamma})
cannot be explained by invoking
any kind of forward scattering disorder.
We are aware of the fact that our explanation
of the constant $\Gamma_0$  in Eq.(\ref{eq:Gamma}) is
based on a plausible argument rather than on a microscopic
calculation. Such a calculation should treat both
forward and backward scattering by impurities and take into
account also
the weak coupling between the chains.

In summary, we have shown that the 
frequency-dependent part of the scattering rate seen 
in the low-frequency conductivity data\cite{Schwartz98}
of the quasi one-dimensional
organic conductors
(TMTSF)$_2$X can be explained by
impurity scattering with small momentum transfers.
The data therefore do not necessarily imply that at small energy scales
the organic conductors become
conventional 
three-dimensional Fermi liquids.
Note that the impurity scattering
mechanism proposed here 
leads for temperatures $T \ll E_F$ to a {\it{temperature-independent}} 
scattering rate.
This can be easily verified from
Eq.(\ref{eq:Mem}), which is also valid at finite
temperatures if we use the finite temperature
density-density correlation function.
Keeping in mind that
electron-electron or electron-phonon scattering
in general lead to temperature-dependent 
scattering rates,
it should be possible to distinguish
the impurity scattering mechanism proposed here
from other mechanisms by measuring 
the temperature-dependence of the scattering rates\cite{Gruener98}.
Note that in several experiments 
on organic conductors a $T^{2}$-behavior
of the resistivity has been observed\cite{Bechgaard80}.
This cannot be explained by invoking forward scattering
by impurities.
If the quadratic frequency-dependence
of the scattering rate seen in the
experiment \cite{Schwartz98} is really due to impurities, then
Eqs.(\ref{eq:Gammares},\ref{eq:disres}) and (\ref{eq:alphares})  
provide a simple way to estimate the
impurity correlation length $\xi$ from
the low-frequency data for the
conductivity.

We would like to thank Ward Beyermann, Martin Dressel, 
George Gr\"{u}ner, and Kurt Sch\"{o}nhammer
for comments on the manuscript.
PK acknowledges  financial supported from the Heisenberg program 
of the DFG.

%

%
%
%
%

\begin{thebibliography}{99}
\bibitem{Throughout}
Throughout this work we shall use units where
$ \hbar =1$.
Although Eqs.(\ref{eq:Gamma}) and (\ref{eq:meff})
seem to fit the data rather well, one should keep
in mind that
the precise determination of the frequency-dependence
of the scattering rate is rather sensitive 
to the normalization and extrapolation of the data.
\bibitem{Comment}
An upper limit for the crossover temperature is given by the hopping
integral $t_{\perp}$ between the chains: $T
{ \raisebox{-0.5ex}{$\; \stackrel{<}{\sim} \;$}} t_{\perp}$.
\bibitem{footnote}
Note that 
Eqs.(\ref{eq:Gammares}) and (\ref{eq:disres})
are the leading behavior for small $\omega$, so that
it would be incorrect to perform a Kramers-Kronig 
analysis on these expressions.
In contrast, the phenomenological Eqs.(\ref{eq:Gamma})
and (\ref{eq:meff}) interpolate from small to large
frequencies.
\bibitem{Gruener98}
We would like to thank George Gr\"{u}ner for pointing this out to us.
\end{thebibliography}
\end{document}